\documentclass[%
reprint,
 amsmath,amssymb,
 aps,
]{revtex4-1}

\usepackage{graphicx}
\usepackage{dcolumn}
\usepackage{bm}

\begin{document}

\preprint{APS/123-QED}

\title{ Cable equation for  general geometry}

\author{Erick J. L\'opez-S\'anchez}
\email{lsej@unam.mx}
\affiliation{Posgrado en Ciencias Naturales e Ingenier\'ia,  Universidad Aut\'onoma Metropolitana, Cuajimalpa. \\
 Vasco de Quiroga 4871, Santa Fe Cuajimalpa,  Ciudad de M\'exico 05300, M\'exico.}
 \author{Juan M. Romero}
\email{jromero@correo.cua.uam.mx}
\affiliation{Departamento de Matem\'aticas Aplicadas y Sistemas,
Universidad Aut\'onoma Metropolitana-Cuajimalpa, Vasco de Quiroga 4871, Santa Fe Cuajimalpa,  Ciudad de M\'exico 05300, M\'exico.}
\date{\today}

\begin{abstract}
The cable equation describes the voltage in  a straight cylindrical cable, this model has been employed to model   electrical potential in dendrites and axons. 
However, sometimes this  equation  might  give incorrect predictions for  some realistic geometries, in particular  when the radius of the cable changes significantly.
Cables with  a non constant radius are important  for some phenomena, for example discrete  swellings along the axons  appear in   neurodegenerative diseases such as  Alzheimer, Parkinson, HIV-associated dementia and Multiple Sclerosis. In this paper, using the Frenet-Serret frame,  we propose a generalized cable equation for  a  general cable geometry. This generalized  equation depends on geometric quantities such as the  curvature and torsion of the cable. We show that when the cable has a constant circular cross-section, the first fundamental form of the cable can be simplified and   the generalized cable equation  depends on  neither the curvature nor  the torsion of the cable.   Additionally, we find an exact solution for an ideal  cable which has a particular variable circular cross-section and zero curvature. For this case we show  that when the cross-section of the cable  increases   the voltage decreases. Inspired in this ideal case, we rewrite the generalized cable equation as a 
diffusion equation with a source term generated by the cable geometry.  This  source term depends on the cable  cross-section area and its derivates. In addition, we study different cables with swelling and provide  their numerical solutions.  
The numerical solutions show that when the cross-section  of the cable has  abrupt changes,  its  voltage is smaller than  the voltage in the  cylindrical cable. Furthermore, these numerical solutions show that the voltage can be affected by  geometrical inhomogeneities on the cable.

\begin{description}
\item[PACS numbers]  87.19.ld, 87.10.Ed, 87.10.Ca, 02.40.Hw
\end{description}
\end{abstract}

\maketitle

\section{Introduction}
Understanding how the brain works is relevant for different  medical and technological applications.   
In this respect it is  central to know  the  electrical brain activity. 
The basic  unit  in the brain is given by neurons, which  transmit electrical signals through axons and receive electrical signals through dendrites. Thus,  it is important understand the electrical behaviour of dendrites and axons. Axons and dendrites can be  described as cables 
with special   properties. It is worth to mentioning that the first  model for an electrical cable was  proposed by Lord Kelvin in the telegraph problem context.  Inspired in the Lord Kelvin's work, different authors have proposed  models to describe dendrites and axons. For example, Rall proposed that a dendrite can be  taken  as a
cable with a circular cross-section and constant diameter $d_{0}$ where the  voltage $V (x,t)$ satisfies
the cable equation
\cite{ralla,rallb,ermentrout,ermentrout2} 
\begin{eqnarray}
c_{M}\frac{\partial V(x,t) }{\partial t}= \frac{d_{0} }{4r_{ L} }  \frac{\partial^{2} V(x,t) }{\partial x^{2}}-  i_{ion},
\label{c1}\qquad
 \end{eqnarray}
here $c_{M}$ denotes the specific membrane capacitance, $r_{L}$ denotes the
longitudinal resistance and $i_{ion}$ is the ionic current  per unit  area  into and out of the cable. 
In the passive cable case, namely when    $i_{ion}=V/r_{M},$ with  $r_{M}$
 the  specific  membrane resistance,  the equation  
(\ref{c1})  is  exactly solved  \cite{ermentrout,ermentrout2,timo}.  \\

The cable equation  has been useful to explain different  phenomena  in dendrites and axons \cite{ermentrout,ermentrout2,timo}.  However,  there are axons and dendrites with different geometry, in particular axons  and dendrites with a variable radius, and the cable  equation  only describes cylindrical cables with a constant radius. Furthermore, there are phenomena where the geometry of the axons and dendrites is relevant. For example, 
axons with a non constant radius   are hallmark features  of   some neurodegenerative diseases. Actually, 
discrete  swellings along the axons  appear in   neurodegenerative diseases such as  Alzheimer, Parkinson, HIV-associated dementia and Multiple Sclerosis \cite{maia,maia1,yin}.
Remarkably, in an extreme  cases, the electrical signal is deleted in the swelling of the axon. Additionally,  theoretical and experimental studies show that dendritic geometry determines the efficacy of voltage propagation \cite{vetter,bird}.
Another example of dendrites  with non trivial geometry is given by spiny dendrites which exhibit  anomalous diffusion \cite{f3}. 
 In addition,  dendrites with varying diameter are
found in synaptic contacts, retina amacrine cells, the cerebellar
dentate nucleus and the lateral vestibular nucleus, as well as
cortical pyramidal and olfactory bulb cells
\cite{kandel,fiala,poznanski1}. Furthermore,  recent theoretical studies emphasize the spatial
variability of dendritic calcium dynamics due to local changes in a
dendrite diameter \cite{tre3}.  For these reasons, it is important to study cables with general geometry.
Some extensions  of the cable model can be seen in \cite{linsay,bedard1,bedard2}.\\

In this paper, using the Frenet-Serret frame,  we propose a cable with general geometry and  construct 
a generalized  cable equation for the voltage in it. This generalized  equation depends  on geometric quantities as  curvature and torsion of the 
cable. For the general case, this  new equation is very complicated. Nevertheless, we show that when the cable has 
 a constant  circular cross-section  the generalized cable  equation  depends  on neither the curvature nor the torsion of the cable. In fact, in this last case  the new cable equation is  equivalent to the cable equation for a straight cylindrical cable. Moreover, we find an exact solution for an ideal cable with a particular variable circular cross-section and zero curvature.  In this case, we show that when radius increases the voltage decreases. 
Inspired in this ideal case, we rewrite the generalized cable equation as a diffusion equation with a source term.
In this diffusion equation the source term and the diffusion coefficient are generated by the cable geometry.
In addition, we provide  numerical solutions to  different cable with swelling.  
The numerical solutions show that when the radius of the cable has  notable changes its  voltage is smaller than  the voltage in the  cylindrical cable. Furthermore, the numerical solutions show that the voltage can be affected by  geometrical inhomogeneities on the cable. 
These   numerical results are  consistent  with the behaviour of the voltage in  focal axonal swellings \cite{maia}.\\

This paper is organized as follows: in the section \ref{se:cg} we propose a cable with a general geometry;  in the section \ref{se:gce} we propose a generalized cable equation; in the section \ref{se:cics} we study a cable with a circular cross-section; in the section \ref{se:rv} we study a particular cable with a  variable radius;  
in the section \ref{se:gep} we study some general properties of the generalized  cable equation; in the section \ref{se:ns} we provide numerical solutions to the  cable equation. Finally, in the section \ref{se:s} a summary is given.

\section{Cable geometry }
\label{se:cg}

It is well known that a three dimensional curve $\vec \gamma$ can be reparametrized with different parameters and its geometric properties are invariant under reparametrizations. For example, the arc length of the curve $\vec \gamma$ is given by 
\begin{eqnarray} 
s=\int_{0}^{x}\sqrt{\frac{d\vec \gamma(\zeta) }{d\zeta }\cdot \frac{d\vec \gamma(\zeta) }{d\zeta }} d\zeta, \label{affine}
\end{eqnarray}
which is invariant under  reparametrization  on $\zeta.$ Notably, the arc length  parameter (\ref{affine}) is a friendly  parameter  to study the  geometric 
properties of a tridimensional curve. For instance, 
 using the arc length parameter (\ref{affine}) we can construct the vectors of the Frenet-Serret  frame \cite{docarmo}
\begin{eqnarray} 
\frac{d \vec \gamma (s)}{ds}=\hat T, \qquad \hat N= \frac{\frac{d \hat T}{ds}  }{ \left| \frac{d \hat T}{ds}\right | }, \qquad 
\hat B=\hat T\times \hat N, 
\end{eqnarray}
where $\hat T$ is the unit vector tangent,  $\hat N$ is the normal unit vector and $\hat B$ is the binormal unit vector to the curve. \\

Furthermore, using the arc length and the Frenet-Serret  frame,   the Frenet-Serret formulas  can be obtained as follow \cite{docarmo}
\begin{eqnarray} 
\frac{d\hat T}{ds}=\kappa \hat N,\quad 
\frac{d\hat N}{ds}=-\kappa \hat T+\tau \hat B,\quad 
\frac{d\hat B}{ds}=-\tau \hat N,
\end{eqnarray}
 here  $\kappa, \tau$ are the curvature and torsion of the curve $\vec \gamma,$ respectively. \\

\begin{figure*}
        \includegraphics[width=0.7\textwidth]{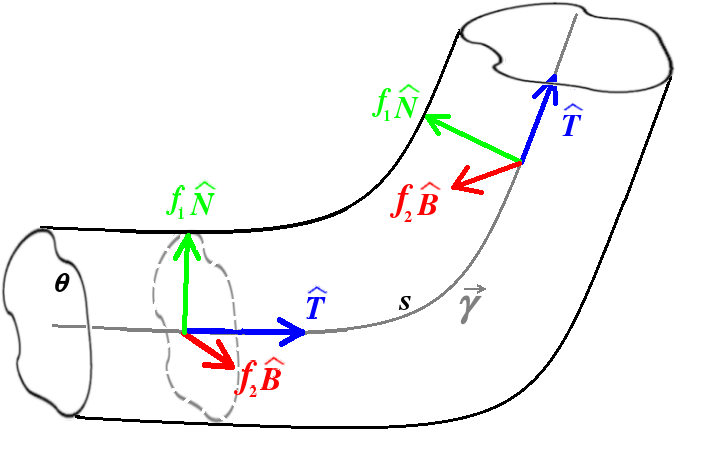}
        \caption{ Cable with general geometry. The vectors  $\hat T, \hat N, \hat B$ are shown in two different points on the curve $\vec \gamma$}\label{tubo}
     \end{figure*}

We can employ the Frenet-Serret  frame to construct a cable model. Actually, 
we can propose  a general cable as the region bounded by the following surface 
\begin{eqnarray}
\vec \Sigma(\theta,s)=\vec \gamma(s)+f_{1}(\theta,s)\hat N(s)+f_{2}(\theta,s) \hat B(s), \label{surface}
\end{eqnarray}
here $\theta$ is an angular variable. Notice that employing the angular coordinate $\theta,$ the functions $f_{1}(\theta,s), f_{2}(\theta,s)$ and the vectors  
$\hat N(s), \hat B(s)$ we are constructing the cable over the curve $\vec \gamma(s).$ For example, 
a  cable with a circular cross-section with  radius $R(s)$ can be parameterized with the functions
\begin{eqnarray}
f_{1}(\theta,s)= R(s)\cos \theta, \quad f_{2}(\theta,s)=R(s)\sin \theta.   \label{circular}
\end{eqnarray}
In the figure \ref{tubo} we can see a representation of the surface (\ref{surface}). \\
  
Some geometric quantities as the area of a surface can be written in terms of the first fundamental form, which  is constructed with  the inner product on the tangent space of a surface   \cite{docarmo}. In the case of  the  surface  (\ref{surface}) we have 
\begin{eqnarray}
 g =
\left( 
\begin{array}{rrrr}
E &  F\\
F & G \\
\end{array}
\right), \label{fff}
\end{eqnarray}
where the first form coefficients are 
\begin{eqnarray}
E&=& \left|\frac{ \partial \vec \Sigma(\theta,s)}{\partial s}\right|^{2} =\left(1-\kappa f_{1}\right)^{2} +\left(\frac{\partial f_{1}}{\partial s} -\tau f_{2} \right)^{2}\nonumber\\
& & +\left(\frac{\partial f_{2}}{\partial s} +\tau f_{1}  \right)^{2},\\
G&=& \left|\frac{ \partial \vec \Sigma(\theta,s)}{\partial \theta}\right|^{2}=  \left( \frac{ \partial f_{1}}{\partial \theta} \right)^{2}+  \left( \frac{ \partial f_{2} }{\partial \theta}\right)^{2}, \\
F&=& \frac{ \partial \vec \Sigma(\theta,s)}{\partial s}\cdot \frac{ \partial \vec \Sigma(s,\theta)}{\partial \theta}= \left( \frac{\partial f_{1} }{\partial s} -\tau f_{2} \right)\frac{\partial f_{1} }{\partial \theta} \nonumber\\
& &+
 \left( \frac{\partial f_{2} }{\partial s} +\tau f_{1} \right)\frac{\partial f_{2} }{\partial \theta}. 
\end{eqnarray}
Using the first fundamental form, an area element of the cable surface (\ref{surface}) can be written as 
\begin{eqnarray}
\Delta A= \left( \int_{0}^{2\pi } d\theta \sqrt{\det g(\theta, s) }\right) \Delta s, \label{area}
\end{eqnarray}
where
\begin{eqnarray}
\det g(\theta,s) &=&
\left[ \left( \frac{\partial f_{1}}{\partial s }\frac{\partial f_{2}}{\partial \theta } -  \frac{\partial f_{2}}{\partial s }\frac{\partial f_{1}}{\partial \theta }\right) 
 -\frac{ \tau}{2} \frac{\partial }{\partial \theta} \left( f_{1}^{2}+f_{2}^{2} \right) \right]^{2} \nonumber \\
& &+\left( 1-\kappa f_{1}\right)^{2}\left[  \left(\frac{\partial f_{1}}{\partial \theta  }\right)^{2} +   \left(\frac{\partial f_{2}}{\partial \theta  }\right)^{2}\right] . 
\end{eqnarray}
Notice, that the area element (\ref{area}) depends on the curvature $\kappa$ and the torsion $\tau$ of the curve $\vec \gamma.$\\

In particular, when the   cable has  a circular cross-section with  radius $R(s),$ namely when the cable is reparametrized with the functions (\ref{circular}),  the  area element of cable surface (\ref{surface}) is given by 
\begin{eqnarray}
 \Delta A& =&R(s) \Delta s \nonumber\\
& & \times  \int_{0}^{2\pi} d\theta \sqrt{ ( 1-\kappa(s) R(s) \cos \theta  )^{2} +\left( \frac{dR(s)}{ds}\right)^{2}}, \label{sa}\qquad 
\end{eqnarray}
while the cable  cross-section area is 
\begin{eqnarray}
 a(s)=\pi R^{2}(s). 
\end{eqnarray}
We can see that  the  area element  (\ref{sa}) does not depend  on the torsion $\tau$ of the curve $\vec \gamma.$ \\

Now,  the  Gaussian curvature characterizes the curvature of a surface \cite{docarmo}.
In particular, if the  cable  (\ref{surface}) has  a circular cross-section with a constant radius $R_{0},$ the 
Gaussian curvature of the surface of this cable is given by
\begin{eqnarray}
K=-\frac{\kappa(s) \cos \theta }{R_{0}\left( 1-\kappa(s) R_{0} \cos \theta \right)}.\label{gaussian}
\end{eqnarray}
Notice that this quantity   is singular  when $\kappa(s) R_{0} \geq 1.$ 
 Because of   the surfaces of  the axons or dendrites reported are smooth surfaces, 
we can suppose that  the Gaussian curvature of  the surfaces of    axons or dendrites does not have singularities. 
Notice that this hypothesis implies the    inequality  $\kappa(s) R_{0} < 1.$ 
  In addition, let  us  remember that the curvature $\kappa(s)$ at a point $P$ of the curve $\vec \gamma(s)$ is defined as the inverse of the radius  of the osculating circle at $P,$ see \cite{spivak}.
Then, if at this   point $P$ the cable radius  is $R(s),$  the radius of  the osculating circle must be larger than $R(s).$ Thus, at the point $P$ the curvature $\kappa(s)$ must be smaller than $R^{-1}(s)$ and the following inequality 
\begin{eqnarray}
\kappa(s) R(s) < 1  \label{condition}
\end{eqnarray}
is satisfied.  In this paper we suppose that the inequality  (\ref{condition}) is always satisfied.\\

Furthermore, notice  that when $R(s)=R_{0}=$constant 
the area element (\ref{sa}) is given by 
\begin{eqnarray}
\Delta A= 2\pi R_{0} \Delta s
\end{eqnarray}
and it depends  on neither the torsion $\tau$  nor  the  curvature $\kappa.$\\

In the next section we will propose a cable equation  when  a cable  is described  by (\ref{surface}).

\section{Cable equation}
\label{se:gce}

In order to propose a cable equation to  a cable with the geometry given by the equation (\ref{surface}), 
we break the curve $\vec \gamma(s)$ into  $n$  pieces. Each piece has a surface area $\Delta A_{i}$ and  a cross sectional area 
$a_{i}$ ($i=0,1,2\cdots n).$ \\

Now, we consider a current flow $I_{long}$ along the cable. Then, if $V(s,t)$ is the membrane potential and $R_{L}$ is the 
resistance of the  cable, from the Ohm's law we have
\begin{eqnarray}
V(s+\Delta s,t)-V(s,t)=-I_{long}(s,t)R_{L}. \label{ohm}
\end{eqnarray}
In addition, due  that the resistance to  a  cable with a cross-sectional area $a(s)$ is \cite{jackson} 
\begin{eqnarray}
R_{L}=\frac{r_{L} \Delta s}{ a(s)},
\end{eqnarray}
where $r_{L}$ is the specific intracellular resistivity,  from the Ohm's law (\ref{ohm}) we obtain 
\begin{eqnarray}
I_{long}&=&-\frac{a(s)}{r_{L}}\left( \frac{V(s+\Delta s,t)-V(s,t)}{\Delta s}\right )\nonumber\\
& \approx&  -\frac{a(s)}{r_{L}}\ \frac{\partial V(s,t)}{\partial s}.\end{eqnarray}
Furthermore, if $C_{M}$ is the membrane capacitance, we get 
\begin{eqnarray}
Q_{cap}=VC_{M},
\end{eqnarray}
which implies 
\begin{eqnarray}
I_{cap}=\frac{dQ}{dt}=C_{M}\frac{ \partial V(s,t)}{\partial t}.\label{capaci}
\end{eqnarray}
Observe   that the capacitance for the cable surface can be taken as 
\begin{eqnarray}
C_{M}= \Delta A(s)c_{M},
\end{eqnarray}
where  $c_{M}$ is  the specific membrane capacitance. Thus, the equation (\ref{capaci}) can be written as 
\begin{eqnarray}
I_{cap}=\Delta A(s)c_{M}\frac{ \partial V(s,t)}{\partial t}.\label{capaci2}
\end{eqnarray}
Moreover, the total ionic current 
that flows across the  membrane is 
\begin{eqnarray}
I_{ion}=(\Delta A )i_{ion},
\end{eqnarray}
where  $i_{ion}$ is the current per unit  area into and out of the cable.\\

Hence, the change in cable current is given by  
\begin{eqnarray}
I_{cap}+I_{ion}=-I_{long}(s+\Delta s,t)+ I_{long}(s,t),
\end{eqnarray}
which implies 
\begin{eqnarray}
& & \Delta A(s)c_{M} \frac{ \partial V(s,t)}{\partial t}+(\Delta A )i_{ion}\nonumber\\
& & \approx \frac{a(s+\Delta s)}{r_{L}} \frac{\partial V(s+\Delta s,t)}{\partial s}
 -\frac{a(s)}{r_{L}}\ \frac{\partial V(s,t)}{\partial s}\nonumber,
\end{eqnarray}
then 
{\small \begin{eqnarray}
\frac{ \partial V(s,t)}{\partial t}+\frac{Êi_{ion}}{c_{M} }\approx\frac{ a(s+\Delta s)  \frac{\partial V(s+\Delta s,t)}{\partial s}
 -a(s)\frac{\partial V(s,t)}{\partial s}}{r_{L}\Delta A(s)c_{M}} .\qquad \label{cap1}
\end{eqnarray} }
Because   of  the  cable surface area is given by the equation (\ref{area}), we obtain 
\begin{eqnarray}
& &\frac{ \partial V(s,t)}{\partial t}+\frac{ i_{ion} }{c_{M}} 
 \approx   \frac{1}{r_{L} c_{M}   \int du \sqrt{\det g(u,s)}} \nonumber\\
 & &
 \left( \frac{ a(s+\Delta s)  \frac{\partial V(s+\Delta s,t)}{\partial s}
 -a(s)\frac{\partial V(s,t)}{\partial s}}{\Delta s } \right) \nonumber.
\end{eqnarray}
Thus, at the limit  $\Delta s \to 0$ we arrive to
\begin{eqnarray}
\frac{\partial V(s,t)}{\partial t}&=&\frac{1}{r_{L} c_{M}  \int_{0}^{2\pi} d\theta \sqrt{\det g(\theta,s) }  } \frac{ \partial }{\partial s} \left (a(s) \frac{\partial  V(s,t)  } {\partial s } \right)\nonumber\\
 & &-   \frac{i_{ion}  }{ c_{M} }.\qquad 
\label{cableg} 
\end{eqnarray}
This  equation  is the cable equation when the cable geometry is given by (\ref{surface}) and it depends on geometric  quantities as the curvature $\kappa$ and torsion $\tau$ of the cable. \\

In the general case, $i_{ion}$ depends on the voltage and the equation (\ref{cableg}) is a non linear differential equation.  
However,  in the  passive cable model we can take 
\begin{eqnarray}
i_{ion}=\frac{V(s,t)}{r_{M}}.
\end{eqnarray}
Therefore,  the cable equation for the passive cable model with the geometry given by (\ref{surface}) is
\begin{eqnarray}
\frac{\partial V(s,t)}{\partial t}&=&\frac{1}{r_{L} c_{M}   \int_{0}^{2\pi} d\theta  \sqrt{\det g(\theta,s) }} \frac{ \partial }{\partial s} \left (a(s) \frac{\partial  V(s,t)  } {\partial s } \right)\nonumber\\
& & -   \frac{V(s,t)  }{ r_{M}c_{M} }.
\label{cableg2} 
\end{eqnarray}
For an infinite cable, the voltage has to satisfy the Dirichlet boundary condition, while for a finite cable the voltage has to satisfy the Neumann boundary  condition \cite{ermentrout}.\\

In the next section we will study some exactly solvable cases of  the equation (\ref{cableg2}).

\section{Circular cross-section}
\label{se:cics}

A cable with  a deformed circular cross-section where the radius depends on the angle $\theta$, namely  $R=R(\theta,s),$ can be modeled with the surface (\ref{surface})
where 
\begin{eqnarray}
f_{1}(\theta,s)=R(\theta,s)\cos \theta, \qquad f_{2}(\theta,s)=R(\theta,s)\sin\theta.  \qquad 
\end{eqnarray}
In this case   the cross-section area is given by
\begin{eqnarray}
a(s)=\frac{1}{2}\int_{0}^{2\pi} R^{2}(\theta,s)d\theta, \label{ihc}
\end{eqnarray}
in addition we obtain 
{\small \begin{eqnarray}
& &  \sqrt{\det g(\theta,s) }=\nonumber \\
&& \Bigg[R^{2}(\theta,s) \left(  \frac{\partial R(\theta,s) }{\partial s} -
\tau \frac{\partial R(\theta,s) }{\partial \theta} \right)^{2}\nonumber\\
& & +\left(1-\kappa(s) R(\theta,s) \cos \theta\right)^{2} \left( R^{2}(\theta,s) + \left( \frac{\partial R(\theta,s) }{\partial \theta}\right)^{2}  \right) \Bigg]^{\frac{1}{2}}.\qquad \label{saa}
\end{eqnarray}}
For this general cable geometry,  to find  solutions of  the cable equation (\ref{cableg2})  is a difficult task. However, for  some cases  this equation can be simplified.
 For instance, when the cable has   a circular cross-section, 
namely when $R(\theta,s)=R(s),$  the function  (\ref{saa}) does not depend on 
the torsion of the cable $\tau.$ Hence,  in this case the cable equation (\ref{cableg2}) becomes
\begin{eqnarray}
 & &\frac{\partial V(s,t)}{\partial t}=\nonumber\\
 & &\frac{\pi \frac{ \partial }{\partial s} \left ( R^{2}(s)  \frac{\partial  V(s,t)  } {\partial s } \right) }{r_{L} c_{M} R(s) \int_{0}^{2\pi} d\theta \sqrt{ ( 1-\kappa(s) R(s) \cos \theta  )^{2} +\left( \frac{dR(s)}{ds}\right)^{2}}  } \nonumber  \\
& & -   \frac{V(s,t)  }{ r_{M} c_{M} }.
\label{cableg1} 
\end{eqnarray}
Notice  that if  the sectional area is  constant, that is $R(s)=R_{0}=$constant,    the equation (\ref{cableg1}) is given by
\begin{eqnarray}
\frac{\partial V(s,t)}{\partial t}=\frac{R_{0}}{2c_{M} r_{L} }  \frac{\partial^{2}  V(s,t)  } {\partial s^{2} }-   \frac{V(s,t)  }{ r_{M} c_{M} }.
\label{cablec} 
\end{eqnarray}
Remarkably, this last  equation  depends on neither the curvature nor  the torsion of the cable.  Furthermore, the equation (\ref{cablec}) is  equivalent to the cable equation for a straight cylindrical cable (\ref{c1}). Observe that the equation (\ref{cablec}) depends on the arc length parameter (\ref{affine}) instead of the lab frame coordinate $x.$ This shows that the natural variables for the voltage are given by geometric quantities of the cable.\\

For an infinite cable, the solution of the  equation (\ref{cablec}) is  
\begin{eqnarray}
 V(s,t)=V_{0}l_{0}\sqrt{ \frac{r_{L}c_{M}}{2\pi R_{0} t}}e^{- \frac{r_{ L}c_{M} s^{2}}{2R_{0} t}}e^{-\frac{t}{r_{M} c_{M}}}, \label{cero}
\end{eqnarray}
where $V_{0}$ is a constant with voltage dimensions  and $l_{0}$ is a constant with length dimensions. Notice that in this case the initial condition 
\begin{eqnarray}
 V(s,0)=V_{0}l_{0}\delta(s)
\end{eqnarray}
is satisfied, where $\delta(s)$ is the Dirac  delta function.  Moreover, using the lab frame coordinate $x$ and the equation (\ref{affine}), the voltage (\ref{cero}) can be written as 
\begin{eqnarray}
 V(x,t)=V_{0}l_{0}\sqrt{ \frac{r_{L}c_{M}}{2\pi R_{0} t}}e^{- \frac{r_{ L}c_{M} \left(\int_{0}^{x}\sqrt{\frac{d\vec \gamma(\zeta) }{d\zeta }\cdot \frac{d\vec \gamma(\zeta) }{d\zeta }} d\zeta \right) ^{2}}{2R_{0} t}}e^{-\frac{t}{r_{M} c_{M}}}.\nonumber 
\end{eqnarray}
Thus, if  a cable has  a constant radius $R_{0},$ the voltage in cable depends on neither  the curvature nor  the torsion.
Furthermore, in this case  voltage  is not different from the voltage for the  straight cylindrical cable with the same radius.\\

\begin{figure*}
        \includegraphics[width=1\textwidth]{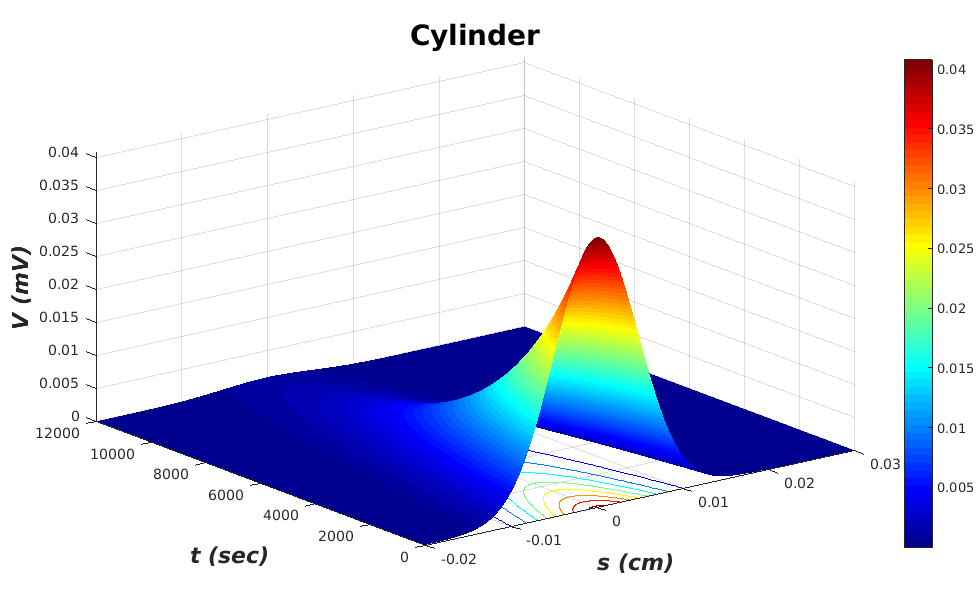}
        \caption{  Voltage for the cylindric cable. Parameter values used for simulations correspond to realistic dendritic parameters as in \cite{poznanski1}: $c_{M} =1mF/cm^2, r_{M} =3000\Omega \cdot cm^2, r_{L} =100\Omega \cdot cm, R_{0} = 10^{-4}cm.$ The initial condition is given by (\ref{inicial}).  }  \label{fig:cyl}
\end{figure*}

The figure \ref{fig:cyl} shows the cable equation solution for a cylindrical cable.  \\

In the following sections we will study cables with  a variable radius.

\section{An exact solution with a variable radius }
\label{se:rv}

Cables with a no constant radius are important for different reasons. 
For example,  discrete swellings along the axons  appear in   
neurodegenerative diseases such as  Alzheimer, Parkinson, HIV-associated dementia and Multiple Sclerosis. 
In particular, in Parkinson disease there are reported  axons with a  diameter  of approximately  $1 \mu m$ with  a swelling with diameter of approximately $5 \mu m,$ see     \cite{Ga99}. In addition,  in  Multiple  Sclerosis there are reported axons with a diameter of approximately   $4 \mu m$
with a  swelling with a diameter of approximately $60 \mu m,$ see \cite{Tr98}. In  Alzheimer disease there are reported axons with a diameter  of approximately  $1.5 \mu m$ with  a swelling train, where the swelling diameter varies between $4 \mu m$ and $10 \mu m,$ see \cite{Xi13, Jo13, Kr13}. 
For HIV-associated dementia, there are reported axons with a diameter of approximately $6\mu m$ and swellings   with a diameter of approximately  $43 \mu m,$
see \cite{Bu87, Bu871,Bu872,Bu873,Bu874}.  Other  sizes of the axonal  swellings can be see in \cite{maia1,Sh98}. \\

When the radius $R(\theta,s)$ is not  a constant, the equation (\ref{cableg1})  is very  complicated. However, some ideal cases can help
us to understand the general case. In this section, we study an ideal case  
which  does not represent a realistic axon or dendrite, but 
it will help us to understand the generalized cable equation (\ref{cableg2}).\\

If  the cable curvature vanishes  and the radius is 
\begin{eqnarray}
R(s)=R_{0}\cosh\left(\frac{s}{R_{0}}\right), \label{radius}
\end{eqnarray}
 the equation (\ref{cableg1}) becomes 
\begin{eqnarray}
\frac{\partial V(s,t)}{\partial t}&=&\frac{R_{0}}{2r_{ L} c_{M}}  \frac{\partial^{2} V(s,t)}{\partial s^{2} } +\frac{1}{r_{L} c_{M}}   \frac{\sinh \left( \frac{s}{R_{0}}\right)}{ \cosh \left( \frac{s}{R_{0}}\right)} \frac{\partial V(s,t)}{\partial s} \nonumber\\ 
& & -   \frac{V(s,t)  }{ r_{M} c_{M} },\nonumber
\end{eqnarray}
which is solved by the voltage
\begin{eqnarray}
 V(s,t)=\frac{V_{0} l^{2}_{0}}{ R(s) }\sqrt{ \frac{r_{L}c_{M}}{2\pi R_{0} t}}e^{- \frac{r_{ L}c_{M} s^{2}}{2R_{0} t}}e^{-t \left( \frac{1}{r_{M} c_{M}}+ \frac{1}{R_{0} r_{L} c_{M} } \right) }. \qquad \label{pradius}
\end{eqnarray}
Notice that when the radius (\ref{radius}) increases the voltage (\ref{pradius}) decreases. \\

   \begin{figure*}
        \includegraphics[width=1\textwidth]{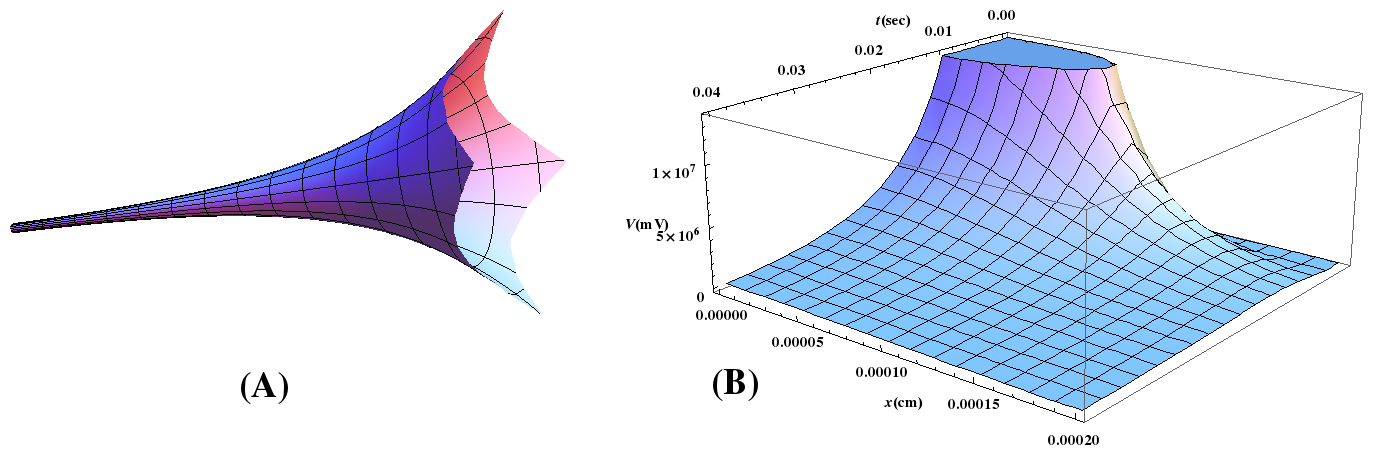}
        \caption{ (A) cable with radius variable (\ref{radius}). (B) Voltage  (\ref{pradius}). Parameter values used for simulations correspond to realistic dendritic parameters as in \cite{poznanski1}: $c_{M} =1mF/cm^2, r_{M} =3000\Omega \cdot cm^2, r_{L} =100\Omega \cdot cm, R_{0} = 10^{-4}cm, V_{0}=1mV, l_{0}=1cm.$  }\label{fig:radv1}
\end{figure*}

In the figure \ref{fig:radv1} we can see the  voltage (\ref{pradius}).

\section{General properties}
\label{se:gep}

Studying  the ideal cable  (\ref{radius})  we learned  that   when the time increases the voltage  (\ref{pradius})  decreases. In addition,  this voltage  depends on $R^{-1}(s)$, namely this voltage depends on 
$(\sqrt{a(s)})^{-1},$ where $a(s)$ is the cable  cross-section area. 
Then, inspired in the ideal  case  (\ref{pradius}), in order to study the equation (\ref{cableg2}),  we  propose  the following voltage 
\begin{eqnarray} 
V(s,t)=\frac{\Psi(s,t)}{\sqrt{a(s)} }. \label{apotentialg}
\end{eqnarray}
Thus,  the equation  (\ref{cableg2}) implies 
\begin{eqnarray} 
\frac{ \partial \Psi(s,t)}{\partial t}= D(s)\frac{ \partial^{2} \Psi(s,t)}{\partial s^{2}}
+\rho(s,t) \qquad \label{dws}
\end{eqnarray}
where 
\begin{eqnarray}
& &D(s)= \frac{a(s)} { r_{L}c_{M}   \int_{0}^{2\pi} d\theta \sqrt{\det g(\theta,s) }},\nonumber\\
& & \rho(s,t)=  \frac{\sqrt{a(s)}}{ r_{L}c_{M}   \int_{0}^{2\pi} d\theta \sqrt{\det g(\theta,s) } } \left( \frac{1}{4}  \Bigg( \frac{ da(s)}{ds} \right)^{2}
\frac{1}{a^{\frac{3}{2}} }\nonumber\\
& & - \frac{1}{2 a^{\frac{1}{2} }(s)} \frac{d^{2} a(s)}{ds^{2} }- \frac{ r_{L}  \int_{0}^{2\pi} d\theta \sqrt{\det g(\theta,s) } }{ r_{M}\sqrt{a(s)} } \Bigg)\Psi(s,t).\qquad \nonumber
\end{eqnarray}
Notice that the equation (\ref{dws}) can be interpreted as a diffusion equation with a source, in this equation the coefficient diffusion and the  source term are generated by the cable geometry.\\

Now, we can see that if we take 
\begin{eqnarray} 
\Psi(s,t)=e^{-Et} \psi(s), \label{apotentialg}
\end{eqnarray}
where $E$ is a constant with (time)$^{-1}$  dimensions, the equation (\ref{dws}) becomes 
\begin{eqnarray} 
 -\frac{\partial^{2} \psi (s)}{\partial s^{2}}+ U(s)\psi(s)=0,\label{asymg}
\end{eqnarray}
here 
\begin{eqnarray}
U(s)& =&  -\frac{  \left(\frac{da(s)}{ds}\right)^{2}}{4a^{2}(s) } 
+\frac{1}{2}\frac{\frac{d^{2}a(s)}{ds^{2}} } {a(s) }  \nonumber\\
 & &-  \frac{ r_{L} c_{M} \left( E-\frac{1}{r_{M} c_{M}} \right)  \int_{0}^{2\pi} d\theta \sqrt{\det g(\theta,s) } }{a(s)} .\label{bpg}
\end{eqnarray}
If  $E=0$ the equation (\ref{asymg}) is  the equilibrium description of the  diffusion  equation (\ref{dws}). 
Moreover, in some cases the Fokker-Planck equation and the diffusion equation can be rewritten as a Schr\"odinger equation \cite{risken}.
In this respect, notice that the equation (\ref{asymg}) can be seen as a Schr\"odinger equation where $U(s)$ is an effective potential,  which is generated by the cable geometry.\\

In particular,   when the radius of the cable is a constant  the effective potential  is   the constant
\begin{eqnarray}
U(s)= U_{0}=-\frac{2 r_{L} c_{M}}{R_{0}} \left( E-\frac{1}{r_{M} c_{M}} \right). 
\end{eqnarray}

When the radius is not a constant, the function $\psi(s)$ is affected by the effective potential (\ref{bpg}).    
For example, if  the cable has  a  circular cross-section with radius $R(s),$ 
the effective potential is 
\begin{eqnarray}
U(s) &=& \frac{1}{R(s)}\Bigg( \frac{1}{2}\frac{d^{2}R(s)}{ds^{2}}-\left( E-\frac{1}{r_{M} c_{M}} \right) \frac{r_{L} c_{M}}{\pi} \times \nonumber\\
 & & \int_{0}^{2\pi } d\theta \sqrt{ \left(1-\kappa(s) R(s) \cos \theta\right)^{2} +\left(\frac{dR(s)}{ds} \right)^{2} }\Bigg) .\qquad \label{effp}
 \end{eqnarray}
From this  last equation  we can see that if the  derivatives of the radius $R(s)$ are small  quantities, then the radius can be approximated by a constant and   the effective potential can be approximated by  a constant   too. 
In this case, the voltage in the cable is similar to the cylindrical cable. Moreover, observe  that the condition (\ref{condition}) implies that $(1-\kappa(s)R(s)\cos \theta)^{2}$ is a small quantity.  Then,  when  $(\frac{dR}{ds})^{2}$ is a large quantity and $(1-\kappa(s)R(s)\cos \theta)^{2}\ll (\frac{dR}{ds})^{2} ,$    the  effective potential (\ref{effp}) depends on neither  the curvature nor the torsion. Thus, in this last case,  cables with similar swelling  have similar voltage.\\

When the cable has  a deformed circular cross-section with   radius   $R=R(\theta,s),$ 
 we should introduce  the equations (\ref{ihc}) and (\ref{saa}) in the 
effective potential (\ref{bpg}). In this case we can see that the effective potential is affected by  derivatives of the radius respect to the angle $\theta.$
Then, geometrical inhomogeneities on a cable  might  be affect the  voltage.\\  

In the next section  we will provide numerical solution for cable with different geometries.\\

\section{Numerical solutions }
\label{se:ns}

In this section   the differential equation (\ref{cableg2})  is solved by using the second order finite differences method for both  spatial coordinate and temporal evolution. The choice of mesh size was made with the usual procedure. First we begin with 1024 points along the s-axis and 50 points in the time. The meshes for s-direction and for the time were refined several times. We stop when no differences in  solutions are obtained in two  successive refinements. The number of spatial and temporal points used in the simulation is shown in  Table \ref{ta:points}.\\

 \begin{table}[hbt]
\begin{center}
\begin{tabular}{|c|c|c|}
\hline
  Refinement     & $n_s$ & $n_t$ \\
\hline
\hline
   First time   & 1024   &  50  \\
 \hline 
   Second time  & 2048   &  80  \\
 \hline 
   Third time   & 4096   &  100  \\
 \hline 
 \hline
\end{tabular}
\caption{\label{ta:points} Number of points, $n_s$ and $n_t$, and number of times of the refinement. }
\end{center}
\end{table}

The system is solved using the Gauss-Seidel iterative method, with a tolerance of $10^{-10}$. Moreover, 
for the initial condition we used a gaussian shape 
\begin{eqnarray}
V(s,0)=\frac{A}{\sqrt{2\pi\sigma}}e^{-\frac{s^2}{2\sigma^2}}, \label{inicial}
\end{eqnarray}
 where $A=0.05 mV\cdot cm^{\frac{1}{2}}$ and $\sigma=100\Delta s.$ In addition  we impose   the   Neumann boundary  conditions \cite{ermentrout} $$\frac{\partial V(s_0,t)}{\partial s} =\frac{\partial V(s_{n_s},t)}{\partial s} = 0. $$\\

\subsection{Cable with sinusoidal swelling }

\begin{figure*}
        \includegraphics[width=1\textwidth]{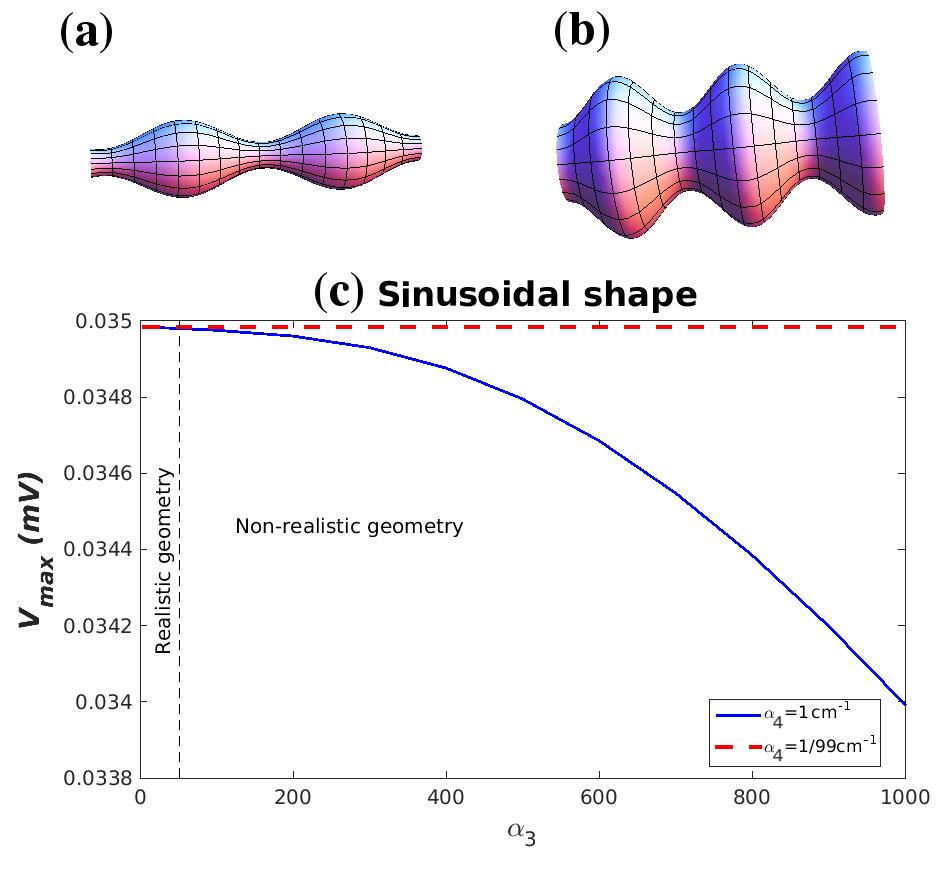}
        \caption{  (a) Cable with geometry (\ref{crnc1}).  (b) Cable with geometry (\ref{crnc2}). (c) Maximum voltage for a cable with radius (\ref{crnc2}) at time $t=1000 sec $ vs the parameter $\alpha_3$. The voltage changes only if $\alpha_{3}>50$, a values for non-realistic cable geometries with swellings. Parameter values used for simulations correspond to realistic dendritic parameters as in \cite{poznanski1}: $c_{M} =1mF/cm^2, r_{M} =3000\Omega \cdot cm^2, r_{L} =100\Omega \cdot cm, R_{0} = 10^{-4}cm.$  The initial condition is given by (\ref{inicial}). }  \label{fig:comparacion}
\end{figure*}

Now we study a cable with curvature  $\kappa=0$ and with the radius  
\begin{eqnarray}
R(s) =  R_{0}\left( 1 + \alpha_{1}\sin\alpha_{2} s \right).\label{crnc1}
 \end{eqnarray}
In the figure \ref{fig:comparacion}(a) we can see a cable with this geometry.
Notice that in this case we should take $\alpha_{1}<1.$ Furthermore,  in order to obtain a realistic cable geometry we should take $\alpha_{2}<1cm^{-1} .$
Then, in this case the derivatives of radius of the cable are small quantities. In fact, for this cable the effective potential is given by 
\begin{eqnarray}
& &U(s) = -\frac{ 1}{1+\alpha_{1} \sin\alpha_{2} s}  \Bigg( \frac{1}{2} \alpha_{1}\alpha^{2}_{2}\sin\alpha_{2}s +\nonumber \\
& &\frac{2}{R_{0}} \left( E-\frac{1}{r_{M} c_{M}} \right) r_{L} c_{M} 
\sqrt{ 1 +R^{2}_{0}\alpha^{2}_{1}\alpha^{2}_{2}\cos^{2} \alpha_{2}s } \Bigg). \label{pcrnc1} \qquad 
 \end{eqnarray}
Notice that for realistic cable geometry the inequality $\alpha_{1}\alpha_{2}\ll1cm^{-1}$ is satisfied and the effective potential is a small quantity. Then,  in this case the voltage in the cable is similar to voltage in the cylindrical cable. \\

A similar result is obtained when the radius of the cable is given by 
\begin{eqnarray}
R(s) =  R_{0}\left( 1 + \alpha_{3}\sin^{2}\alpha_{4} s \right).\label{crnc2}
 \end{eqnarray}
In the figure \ref{fig:comparacion}(b) we can see a cable with this geometry.\\

Notice that in this case $\alpha_{3}$ can be bigger than $1,$ and   for a realistic cable geometry 
the condition $\alpha_{3}\alpha_{4}\ll1$ cm$^{-1}$ should be satisfied. It can be shown that in this  case   the effective potential (\ref{effp}) is a small quantity and then the voltage in the cable is also not different from the  voltage for the cylindrical cable.\\

In the figure \ref{fig:comparacion}(c) we plot the maximum voltage as a function of $\alpha_{3}$ for  two different values of $\alpha_{4}.$
We can see that the voltage to cable with a realistic geometry is not different from the voltage for  a cylindrical cable. \\

\subsection{Cable with  Gaussian swelling}

\begin{figure*}
        \includegraphics[width=1\textwidth]{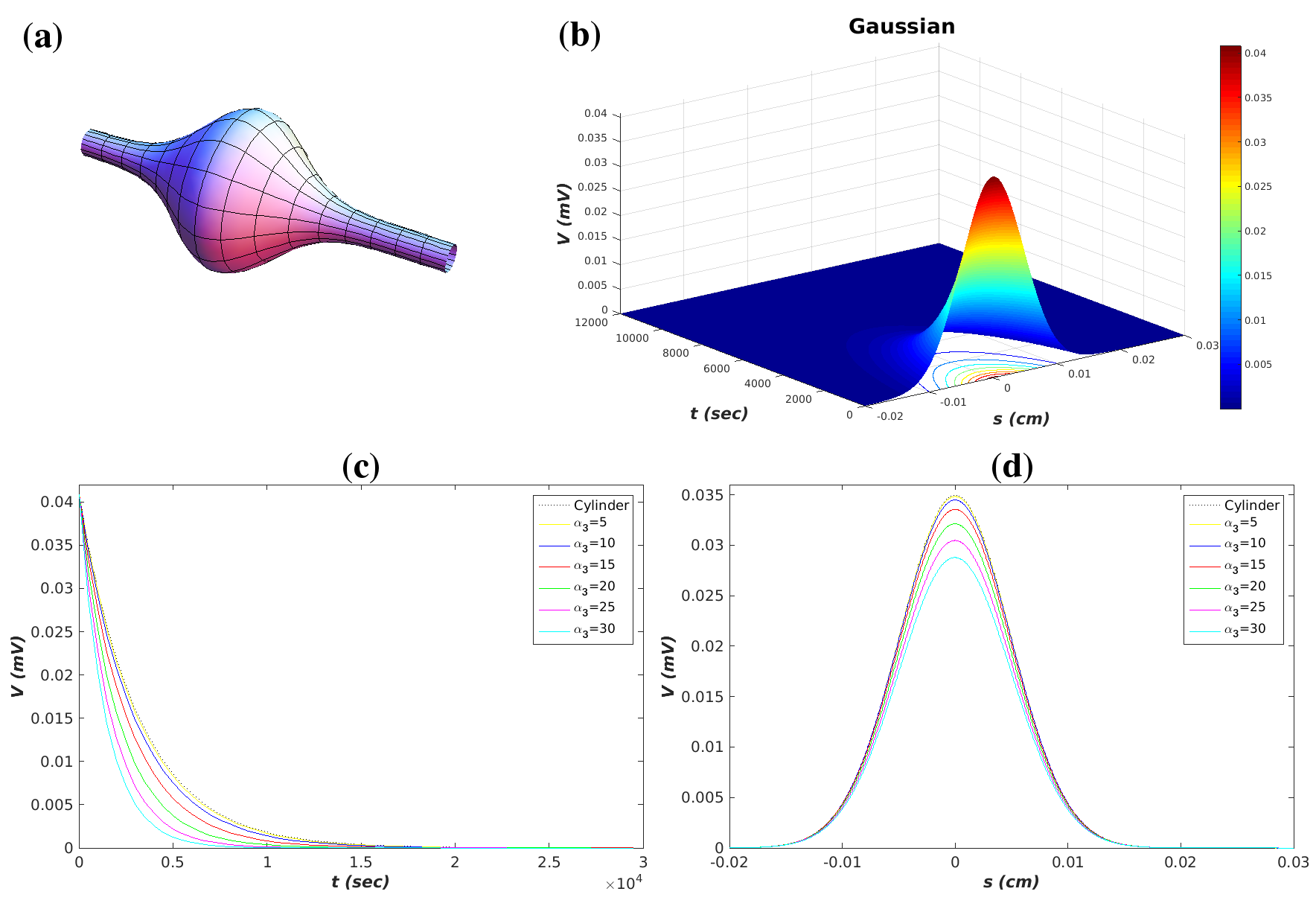}
        \caption{  (a) Cable with geometry (\ref{radius1}). (b)  Voltage for the cable with a gaussian swelling, radius (\ref{radius1}). (c) Voltage vs $t$ in $s=0$ with different values for $\alpha_3$. (d) Voltage vs $s$ at time $t=1000sec$ for different values of $\alpha_3.$ Parameter values used for simulations correspond to realistic dendritic parameters as in \cite{poznanski1}: $c_{M} =1mF/cm^2, r_{M} =3000\Omega \cdot cm^2, r_{L} =100\Omega \cdot cm, R_{0} = 10^{-4}cm.$   The initial condition is given by (\ref{inicial}). }\label{fig:gaussian}
\end{figure*}

Now, we study  a cable with the radius 
\begin{eqnarray}
R(s) =  R_{0}\left( 1 + \alpha_{3} e^{-\alpha_{5}(s - \alpha_{6})^{2}}  \right). \label{radius1}
 \end{eqnarray}
In the figure \ref{fig:gaussian}(a) can see a cable with this geometry.  
 The numerical solution to  the voltage can be seen in the figure \ref{fig:gaussian}(b).  In this figure  we can see that 
 the voltage decreases faster than the voltage to the cylindrical cable. Notice that for this geometry   when the  height   of the swelling (\ref{radius1}) increases the voltage of 
 the cable decreases.  In fact, for a big  $\alpha_{3}$ parameter the voltage can be blocked. 
 
\subsection{Cable with Gaussian swellings }

\begin{figure*}
        \includegraphics[width=1\textwidth]{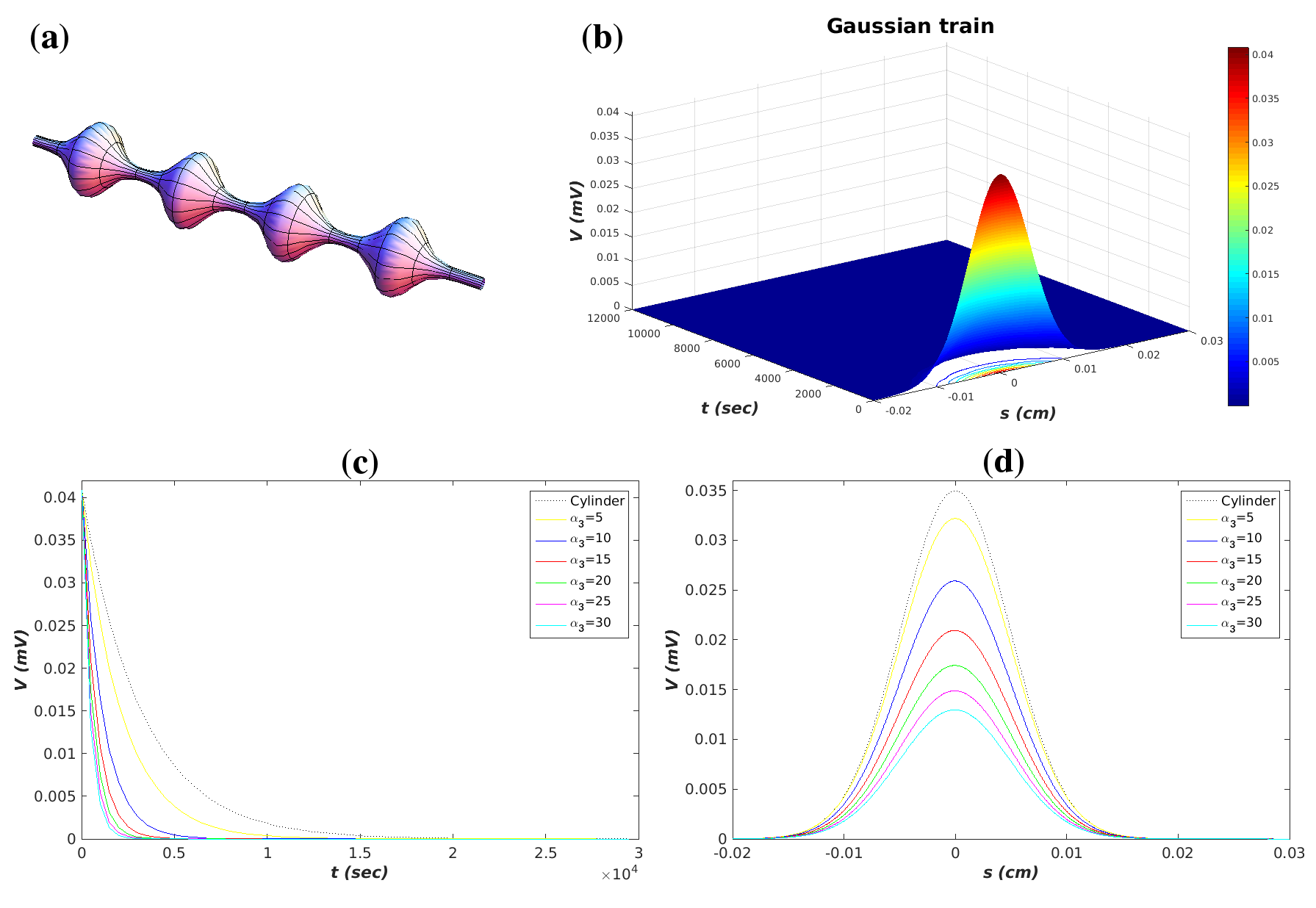}
        \caption{   (a) Cable with geometry (\ref{radius2}). (b)  Voltage for the cable with a gaussian train swellings, radius (\ref{radius2}). (c) Voltage vs $t$ in $s=0$ with different values for $\alpha_3$. (d) Voltage vs $s$ at time $t=1000sec$ for different values of $\alpha_3.$ Parameter values used for simulations correspond to realistic dendritic parameters as in \cite{poznanski1}: $c_{M} =1mF/cm^2, r_{M} =3000\Omega \cdot cm^2, r_{L} =100\Omega \cdot cm, R_{0} = 10^{-4}cm.$  The initial condition is given by (\ref{inicial}).  }\label{fig:gaussiant}
\end{figure*}

In this section we study the cable with the following radius 
\begin{eqnarray}
R(s) &=&  R_{0}\Bigg (1 + \alpha_{3} e^{-\alpha_{5}s^{2}}  + \alpha_{3} e^{-\alpha_{5} (s - \alpha_{6})^{2}}\nonumber\\
& &
+\alpha_{3} e^{-\alpha_{5}(s - 2\alpha_{6})^{2}}  + \alpha_{3} e^{-\alpha_{5} (s - 3\alpha_{6})^{2}} \Bigg).\label{radius2}
 \end{eqnarray}
 A cable with this geometry can be seen in the figure \ref{fig:gaussiant}(a) and  the numerical solution to the voltage in this cable can be seen in the figure \ref{fig:gaussiant}(b).  In this case we can see that  the voltage decreases faster than the voltage to the cylindrical cable.  Moreover, notice that for this geometry   when the  height   of the swelling (\ref{radius2}) increases the voltage of  the cable decreases. Observe that the voltage in this cable decreases more than the voltage in the cable with radius (\ref{radius1}).   Actually, if  $\alpha_{3}$ is a big parameter the voltage can be blocked.  

\subsection{Amorphous swelling }

\begin{figure*}
        \includegraphics[width=1\textwidth]{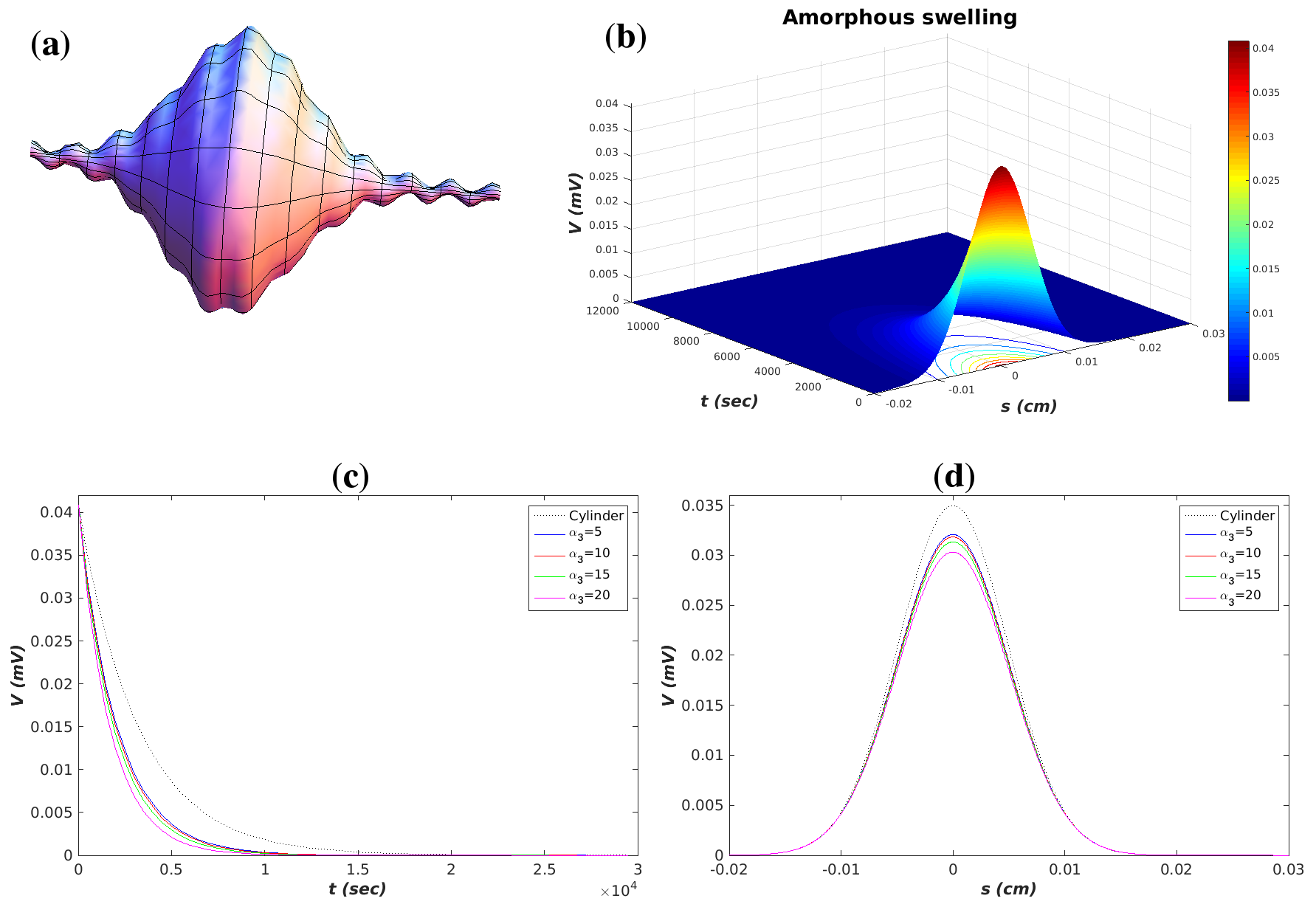}
        \caption{  (a) Cable with geometry (\ref{amorphusc2}). (b)  Voltage for the cable with the radius (\ref{amorphusc2}).  (c) Voltage vs $t$ in $s=0$ with different values for $\alpha_3$. (d) Voltage vs $s$ at time $t=1000sec$ for different values of $\alpha_3.$ Parameter values used for simulations correspond to realistic dendritic parameters as in \cite{poznanski1}: $c_{M} =1mF/cm^2, r_{M} =3000\Omega \cdot cm^2, r_{L} =100\Omega \cdot cm, R_{0} = 10^{-4}cm.$ The initial condition is given by (\ref{inicial}). }\label{fig:gaussianamo1}
\end{figure*}

In the literature there are reported axons with amorphous swelling \cite{maia}. 
For an amorphous cable we can propose the following radius 
\begin{eqnarray}
R(\theta,s) = R_{0}\left(1 +  \alpha_{3}e^{-\alpha_{5}(s - \alpha_{6}) ^{2} } + \alpha_{7} \sin\theta  \cos \alpha_{8}s \right). \label{amorphusc2}\qquad 
\end{eqnarray}
A cable with this geometry can be seen in the figure \ref{fig:gaussianamo1}(a) and  the numerical solution for the voltage in this cable can be seen in the figure 
\ref{fig:gaussianamo1}(b), (c), (d).  In this case we can see that the voltage decreases faster than the voltage to the cylindrical cable.  In addition, observe that the voltage in this cable decreases more than the voltage in the cable with radius (\ref{radius1}). Thus, geometric  inhomogeneities in a cable  can change   its   voltage in it. \\

Axons with swellings  are hallmark features  of   some neurodegenerative diseases such as  Alzheimer, Parkinson, HIV-associated dementia and Multiple Sclerosis \cite{maia,maia1,yin}.
The numerical results of this section indicate us when   the  cable geometry is important  to the voltage propagation. These results show that when the derivatives of the cable radius are slowly changing  functions, namely when  $ (\frac{dR}{ds})^{2} \ll (1-\kappa(s)R(s)\cos \theta)^{2}$ and
 $\frac{d^{2} R(s)} {ds^{2} } \ll \frac{r_{L}}{r_{M}},$  the voltage is similar to voltage in a straight cylindrical cable. However, when  these derivatives  
change  significantly   the voltage cable   is reduced and,  in a extreme case, it  can be blocked. Moreover, these  numerical results show that   geometric  inhomogeneities in a cable  can affect   the  voltage. \\

\section{Summary}
\label{se:s}

In this paper, using the Frenet-Serret frame,  we proposed a cable with general geometry and  construct 
a generalized  cable equation to the voltage in it. This generalized  equation depends  on geometric quantities as  curvature and torsion of the 
cable. For the general case, this  equation is very complicated to obtain exact solutions. However,  when the cable has 
 a constant circular cross-section,  we showed that  the equation depends on neither the curvature nor the torsion of the cable. In fact, in this last case  the new  equation is  equivalent to the cable equation for a straight cylindrical cable, where the voltage depends
on the arc length parameter.  This shows that the natural variables for the voltage are given 
by the cable  geometric quantities.    
Additionally, we found an exact solution 
for an ideal   cable with  a particular non constant circular cross-section and  zero curvature.  In this last case, we show that when the radius increases the voltage decreases. 
Inspired in this ideal case, 
we rewrote the generalized cable equation   as a diffusion equation with a source term.
In this diffusion equation the source term and the diffusion coefficient are generated by the cable geometry.
Furthermore, we provided  numerical solutions to the new cable equation to  different cable with swelling.  
These solutions  show that when the derivatives of the cable radius are slowly changing  functions the voltage is similar to voltage in a straight cylindrical cable. However, when   these derivatives  
change  significantly   the voltage cable   is reduced and,  in a extreme case, it  can be blocked. Moreover, these  numerical results show that   geometric  inhomogeneities in a cable  can change   its   voltage in it. The results of  this paper might help us to understand the the behaviour of the voltage in  focal axonal swellings which appears in some neurodegenerative diseases such as  Alzheimer, Parkinson, HIV-associated dementia and Multiple Sclerosis \cite{maia}.\\

In this work we did not study the active case, but we can argue that when the cable has a constant circular cross-section 
the active case is not different  from   the usual  active cylindrical cable. In  a future work we will study the active  cable case and some important cable geometries, for instance the spiny cable geometry.\\
  
\begin{acknowledgments}
This work was supported in part by  CONACyT-SEP 47510318 (J.M.R.).  We are grateful to referees for providing valuable
comments. \\


\end{acknowledgments}


\begin{thebibliography}{00}




\bibitem{ralla}
W. Rall, Cable Theory for Dendritic Neurons,  in {\it Methods in
Neuronal Modeling: From Synapses to Networks,} eds. C. Koch and I.
Segev (MIT Press, 1989) Chap. 2.

\bibitem{rallb}
W. Rall, Theoretical significance of dendritic trees for neuronal
input-output relations, in {\it Neural Theory and Modeling,} ed.  R. F. Reiss,
(Sanford University Press, Stanford, 1964) p. 73.



\bibitem{ermentrout}
G. B.  Ermentrout and  D. H. Terman, {\it  Mathematical foundation of
neuroscience,} Springer, London,  2010 .

\bibitem{ermentrout2}
P. C. Bressloff, {\it Waves in Neural Media: From Single Neurons to
Neural Fields,} Springer, London,  2014.


\bibitem{timo}
Q. Caudron, S. R. Donnelly, S. P. C. Brand and  Y. Timofeeva,   
 J. Math. Neurosci. {\bf 2 (1)}, 11 (2012).
 

\bibitem{maia}
P. D. Maia,  M.A. Hemphill,  B.  Zehnder,  C. Zhang, K. K. Parker,  J. N.  Kutz, {\it Diagnostic tools for evaluating the impact of Focal Axonal Swellings arising in neurodegenerative diseases and/or traumatic brain injury,}  
Journal of Neuroscience Methods   {\bf 253},  233  (2015).  


\bibitem{maia1}
M. H. Magdesian, F. S. Sanchez,  M. Lopez, P. Thostrup, N.  Durisic, W.  Belkaid, D.  Liazoghli, P.  Gr\"utter and D.R. Colman, 
{\it Atomic Force Microscopy Reveals Important Differences in Axonal
Resistance to Injury,}  Biophysical Journal   {\bf 103},    405 (2012).

\bibitem{yin}
T. C. Yin, J. R. Voorhees, R. M. Genova, K. C. Davis, A.  M. Madison, J. K. Britt, C. J. Cintr\'on-P\'erez, L.  McDaniel, M. M. Harper, A. A. Pieper,  {\it Acute Axonal Degeneration Drives Development of Cognitive, Motor, and Visual Deficits after Blast-Mediated Traumatic Brain Injury in Mice,}
eNeuron   {\bf 3} (5)  e0220-16 (2016).   


\bibitem{vetter}
P. Vetter, A. Roth and M. Hausser,
 {\it Propagation of action potentials in dendrites depends on dendritic morphology,}
Neurophysiol {\bf 85}, 926 (2001).


\bibitem{bird}
A. D. Bird and H. Cuntz, {\it Optimal current transfer in dendrites,}
PLoS Comput Biol.  {\bf 12} (5): e1004897 (2016). 



\bibitem{f3}
F. Santamaria, S. Wils, E. De Schutter and G. J. Augustine,  
Neuron {\bf 52}, 4  (2006).


 \bibitem{kandel}
E. R. Kandel,  J. H.  Schwartz and T. N. Jessell,   {\it Principles of Neural
Science}, McGraw-Hill, New York,  2000.

\bibitem{fiala}
J. C. Fiala and K. M. Harris,   Dendrite Structure,  in  {\it  Dendrites,} eds.
G. Stuart, N. Spruston, M. Häusser, (Oxford University Press,  1999) p.1.


\bibitem{poznanski1}
R. R. Poznanski, Bull. Math. Biol. {\bf 54},  905  (1992).



\bibitem{tre3}
H. Anwar, C. J. Roome,  H. Nedelescu, W. Chen, B. Kuhn and  E. De Schutter, 
 Front. Cell. Neurosci. {\bf 8},  168 (2014). 



\bibitem{linsay}
K. A. Lindsay, J. R. Rosenberg and G. Tucker, {\it From Maxwell's equations to the cable equation and beyond,}
Progress in Biophysics and Molecular Biology  {\bf 85},  71 (2004).

\bibitem{bedard1}
C. Bed\'ard and A. Destexhe, 
{\it Generalized cable theory for neurons in complex and heterogeneous media,}
Phys. Rev. E {\bf 88}, 022709  (2013).




\bibitem{bedard2}
C. Bed\'ard and A. Destexhe, 
{\it Generalized cable formalism to calculate the magnetic field of single neurons and neuronal populations,}
Phys. Rev. E {\bf 90}, 042723   (2014)





\bibitem{docarmo}
M. P. Do Carmo, {\it Differential Geometry of Curves and Surfaces,} Dover,  NewYork (2016). 



\bibitem{spivak}
M. Spivak, {\it A Comprehensive Introduction to Differential Geometry,} vol. II,  Publish or Perish (1975). 


\bibitem{jackson}
W. Greiner, {\it Classical Electrodynamics,} Springer-Verlag, NewYork (1991). 





\bibitem{Ga99} J. E. Galvin, K. Uryu, V.  M.-Y. Lee  and J.  Q. Trojanowski,  {\it Axon pathology in Parkinson'\ s disease and Lewy body dementia hippocampus contains $\alpha$-, $\beta$-, and $\gamma$-synuclein,}  Proceedings of the National Academy of Sciences of the United States of America, Proceedings of the National Academy of Sciences {\bf 96} (23), 13450 (1999).


\bibitem{Tr98} B. D. Trapp, J.  Peterson, R. M. Ransohoff, R. Rudick, S. M\"{o}rk  and L.  B\"{o}, {\it Axonal Transection In The Lesions Of Multiple Sclerosis,}   The New England Journal of Medicine  {\bf 338} (5), 278 (1998) .


\bibitem{Jo13} V.  E.  Johnson, W. Stewart and D. H. Smith, {\it Axonal Pathology in Traumatic Brain Injury,}  Experimental  Neurology  {\bf 246}, 35 (2013).

\bibitem{Kr13} D. Krstic and I.  Knuesel, {\it Deciphering the mechanism underlying late-onset Alzheimer disease,}  Nature Reviews Neuroscience    {\bf 9},  25 (2013). 




\bibitem{Xi13} H.  Xie, J.  Guan,  L. A. Borrelli,  J. Xu,  A. Serrano-Pozo  and B.  J. Bacskai,  {\it Mitochondrial Alterations near Amyloid Plaques in an Alzheimer's Disease Mouse Model,}  The Journal of Neuroscience {\bf 33} (43),  17042 (2013).






\bibitem{Bu87} H. Budka, G. Costanzi, S. Cristina, A. Leehi, C. Parravicini, R. Trabattoni, and L. Vago, {\it Brain pathology induced by infection with the human immunodeficiency virus (HIV),}    Acta Neuropathologica  {\bf 75 }, 185 (1987).

\bibitem{Bu871} F. Gray,  F. Chr\'etien,  A. V. Vallat-Decouvelaere and 
F. Scaravilli,  {\it The Changing Pattern of HIV Neuropathology in the HAART Era,} 
Journal of Neuropathology and Experimental Neurology  {\bf 62} (5), 429 (2003).


\bibitem{Bu872}  M.  Kaul, G. A. Garden and  S.  A. Lipton, {\it Pathways to neuronal injury and apoptosis in HIV-associated dementia,} 
Nature {\bf  410},  988 (2001).


\bibitem{Bu873} F. Raja, F. E. Sherriff,  C. S. Morris,  L. R. Bridges and M. M. Esiri,   {\it Cerebral white matter damage in HIV infection demonstrated using $\beta$-amyloid precursor protein immunoreactivity, }  Acta  Neuropathologica {\bf 93}, 184 (1997).



\bibitem{Bu874} R. Ellis, D. Langford and E.  Masliah,  {\it HIV and antiretroviral therapy in therain: neuronal injury and repair,} Nature Reviews Neuroscience  {\bf 8},  33  (2007).



\bibitem{Sh98}  G. M. G.  Shepherd  and K. M. Harris, {\it Three-Dimensional Structure and Composition of CA3$\rightarrow$CA1 Axons in Rat Hippocampal Slices: Implications for Presynaptic Connectivity and Compartmentalization,}  The Journal of Neuroscience  {\bf 18} (20),  8300 (1980).





\bibitem{risken}
H. Risken, {\it The Fokker-Planck Equation, Methods of Solution and Applications,} Springer-Verlag, NewYork (1989). 


\end{thebibliography}
\end{document}